\documentclass[10pt]{article}
\usepackage{amsmath}
\usepackage{amssymb}
\usepackage{graphicx}
\usepackage{cite}
\usepackage{color}
\usepackage[labelfont=bf,labelsep=period,justification=raggedright]{caption}
\makeatletter
\renewcommand{\@biblabel}[1]{\quad#1.}
\makeatother

\date{}

\begin{document}

\begin{flushleft}
{\Large
\textbf{Heterogeneous aspirations promote cooperation in the prisoner's dilemma game}
}\sffamily
\\[3mm]
\textbf{Matja{\v z} Perc$^{1,\ast}$ and
 Zhen Wang$^{2,\dagger}$}
\\[2mm]
{\bf 1} Faculty of Natural Sciences and Mathematics, University of Maribor, Slovenia\\
{\bf 2} School of Physics, Nankai University, Tianjin, China

$^{\ast}$matjaz.perc@uni-mb.si [www.matjazperc.com]; $^{\dagger}$zhenwang@mail.nankai.edu.cn
\end{flushleft}
\sffamily
\section*{Abstract}
To be the fittest is central to proliferation in evolutionary games. Individuals thus adopt the strategies of better performing players in the hope of successful reproduction. In structured populations the array of those that are eligible to act as strategy sources is bounded to the immediate neighbors of each individual. But which one of these strategy sources should potentially be copied? Previous research dealt with this question either by selecting the fittest or by selecting one player uniformly at random. Here we introduce a parameter $u$ that interpolates between these two extreme options. Setting $u$ equal to zero returns the random selection of the opponent, while positive $u$ favor the fitter players. In addition, we divide the population into two groups. Players from group $A$ select their opponents as dictated by the parameter $u$, while players from group $B$ do so randomly irrespective of $u$. We denote the fraction of players contained in groups $A$ and $B$ by $v$ and $1-v$, respectively. The two parameters $u$ and $v$ allow us to analyze in detail how aspirations in the context of the prisoner's dilemma game influence the evolution of cooperation. We find that for sufficiently positive values of $u$ there exist a robust intermediate $v \approx 0.5$ for which cooperation thrives best. The robustness of this observation is tested against different levels of uncertainty in the strategy adoption process $K$ and for different interaction networks. We also provide complete phase diagrams depicting the dependence of the impact of $u$ and $v$ for different values of $K$, and contrast the validity of our conclusions by means of an alternative model where individual aspiration levels are subject to evolution as well. Our study indicates that heterogeneity in aspirations may be key for the sustainability of cooperation in structured populations.

\section*{Introduction}
Understanding the evolution of cooperation among selfish individuals in human and animal societies remains a grand challenge across disciplines. Evolutionary games are employed frequently as the theoretical framework of choice in order to interpret the emergence and survival of cooperative behavior \cite{hofbauer_98, nowak_06, maynard_82, roca_plr09, hauert_ajp05, szabo_pr07}. The prisoner's dilemma game, in particular, has attracted considerable interest \cite{nowak_amc89, moyano_jtb09, doebeli_el05, szabo_pre07, szolnoki_pre09} as the essential yet minimalist example of a social dilemma. In the original two-person one-shot game the two players have two strategies to choose from (cooperation and defection), and their payoffs depend on the simultaneous decision of both. If they choose to cooperate they will receive the highest collective payoff, which will be shared equally among them. Mutual defection, on the other hand, yields the lowest collective payoff. Yet to defect is tempting because it yields a higher individual payoff regardless of the opponent's decision. It is thus frequently so that both players choose not to cooperate, thus procreating the inevitable social dilemma. In reality, however, interactions may be repeated and the reputation of players compromised \cite{milinski_n02, fehr_n04}. Additionally, individuals may alter with whom they interact \cite{van-segbroeck_prl09}, and different behaviors may be expressed
when participants in a social interaction occupy different roles \cite{maynardsmith_ab76, selten_jtb80, hammerstein_ab81, marshall_jtb09}. Such and similar considerations have been very successful in elucidating why the unadorned scenario of total defection is often at odds with reality \cite{milinski_n87}, where it is clear that both humans and animals cooperate to achieve what would be impossible by means of isolated efforts. Mechanisms supporting cooperation identified thus far include kin selection \cite{hamilton_wd_jtb64} as well as many others \cite{nowak_s06, szabo_pr07, schuster_jbp08, perc_bs10}, and there is progress in place aimed at unifying some of these approaches \cite{lehmann_jeb06, marshall_bes10}.

Probably the most vibrant of all in recent years have been advances building on the seminal paper by Nowak and May \cite{nowak_n92b}, who showed that spatial structure may sustain cooperation without the aid of additional mechanisms or strategic complexity. Although in part anticipated by Hamilton's comments on viscous populations \cite{hamilton_wd_jtb64}, it was fascinating to discover that structured populations, including complex and social networks \cite{albert_rmp02, newman_siamr03, boccaletti_pr06}, provide an optimal playground for the pursuit of cooperation. Notably, a simple rule for the evolution of cooperation on graphs and social networks is that natural selection favors cooperation if $b/c>k$, where $b$ is the benefit of the altruistic act, $c$ is its cost, while $k$ is the average number of neighbors \cite{ohtsuki_n06}. This is similar to Hamilton's rule stating that $c/b$ should be larger than the coefficient of genetic relatedness between individuals \cite{hamilton_wd_jtb64}. In fact, on graphs and social networks the evolution of altruism can thus be fully explained by the inclusive fitness theory since the population is structured such that interactions are between genetic relatives on average \cite{lehmann_jeb07, grafen_jeb07, marshall_bes10}.

According to the "best takes all" rule \cite{nowak_ijbc93, nowak_ijbc94} players are allowed to adopt the strategy of one of their neighbors, provided its payoff is higher than that from the other neighbors as well as from the player aspiring to improve by changing its strategy. Based on this relatively simple setup, it was shown that on a square lattice cooperators form compact clusters and so protect themselves against being exploited by defectors. The "best takes all" strategy adoption rule is, however, just one of the many possible alternatives that were considered in the past. Other examples include the birth-death and imitation rule \cite{ohtsuki_prslb06}, the proportional imitation rule \cite{schlag_jet98}, the reinforcement learning adoption rule \cite{wang_s_ploso08}, or the Fermi-function based strategy adoption rule \cite{szabo_pre98}. The latter received substantial attention, particularly in the physics community, for its compatibility with the Monte Carlo simulation procedure and the straightforward adjustment of the level of uncertainty governing the strategy adoptions $K$. However, with this rule the potential donor of the new strategy is selected uniformly at random from all the neighbors. This is somewhat untrue to what can be observed in reality, where in fact individuals typically aspire to their most successful neighbors rather than just somebody random. In this sense the "best takes all" rule seems more appropriate, although it fails to account for errors in judgment, uncertainty, external factors, and other disturbances that may vitally affect how we evaluate and see our co-players. Here we therefore propose a simple tunable function that interpolates between the "best takes all" and the random selection of a neighbor in a smooth fashion by means of a single parameter $u$. In this sense the parameter $u$ acts as an aspiration parameter, determining to what degree neighbors with a higher payoff will be considered more likely as potential strategy sources than other (randomly selected) neighbors.

Aiming to further disentangle the role of aspirations, we also consider two types of players, denoted by type $A$ and $B$, respectively. While players of type $A$ conform to the aspirations imposed by the value of the aspiration parameter $u$, type $B$ players choose whom to potentially copy uniformly at random irrespective of $u$. We denote the fraction of type $A$ and $B$ players by $v$ and $1-v$, respectively. This additional division of players into two groups is motivated by the overwhelming evidence indicating that heterogeneity, almost irrespective of its origin, promotes cooperative actions. Most notably associated with this statement are complex networks, including small-world networks \cite{abramson_pre01, ren_pre07, chen_xj_pre08}, random regular graphs \cite{hauert_ajp05, vukov_pre06}, scale-free networks \cite{santos_prl05, santos_prslb06, poncela_njp07, assenza_pre08, szolnoki_pa08, perc_njp09}, as well as adaptive and growing networks \cite{zimmermann_pre04, pacheco_prl06, szolnoki_epl08, poncela_ploso08, szolnoki_epl09, poncela_njp09, szolnoki_njp09, traulsen_plos10}. Furthermore, we follow the work by McNamara \textit{et al.} on the coevolution of choosiness and cooperation \cite{mcnamara_n08}, in particular by omitting the separation of the population on two types of players and introducing the heterogeneity by means of normally distributed individual aspiration levels that are then also subject to evolution.

At present, we thus investigate how aspirations on an individual level affect the evolution of cooperation. Having something to aspire to is crucial for progress and betterment. But how high should we set our goals? Should our role models be only overachievers and sports heroes, or is it perhaps better to aspire to achieving somewhat more modest goals? Here we address these questions in the context of the evolutionary prisoner's dilemma game and determine just how strong and how widespread aspirations should be for cooperation to thrive best. As we will show, a strong drive to excellence in the majority of the population may in fact act detrimental on the evolution of cooperation, while on the other hand, properly spread and heterogeneous aspirations may be just the key to fully eliminating the defectors. We will show that this holds irrespective of the structure of the underlying interaction network, as well as irrespective of the level of uncertainty by strategy adoptions $K$. In addition, the presented results will be contrasted with the output of a simple coevolutionary model, where individual aspirations will also be subject to evolution by means of natural selection. We will conclude that appropriately tuned aspirations may be seen as a universally applicable promoter of cooperation, which will hopefully inspire new studies along this line of research.

\section*{Results}
Depending on the interaction network, the strategy adoption rule and other simulation details (see \textit{e.g.} \cite{hauert_ijbc02, szabo_pr07, roca_plr09, perc_bs10}), there always exists a critical cost-to-benefit ratio $r=r_c$ at which cooperators in the prisoner's dilemma die out. This is directly related to Hamilton's rule stating that natural selection favors cooperation if $c/b$ is larger than the coefficient of genetic relatedness between individuals \cite{hamilton_wd_jtb64}. If the aspiration parameter $u=0$ (note that then the division of players to those of type $A$ and those of type $B$ is irrelevant), $K=0.1$, and the interaction network is a square lattice, then, in our case, $r_c=0.022$. In what follows, we will typically set $r$ slightly below this threshold to $0.02$ and examine how different values of $u$, $v$, $K$, as well as different interaction networks influence the outcome of the prisoner's dilemma game.

\begin{figure}[!ht]
\includegraphics[width=15.75cm]{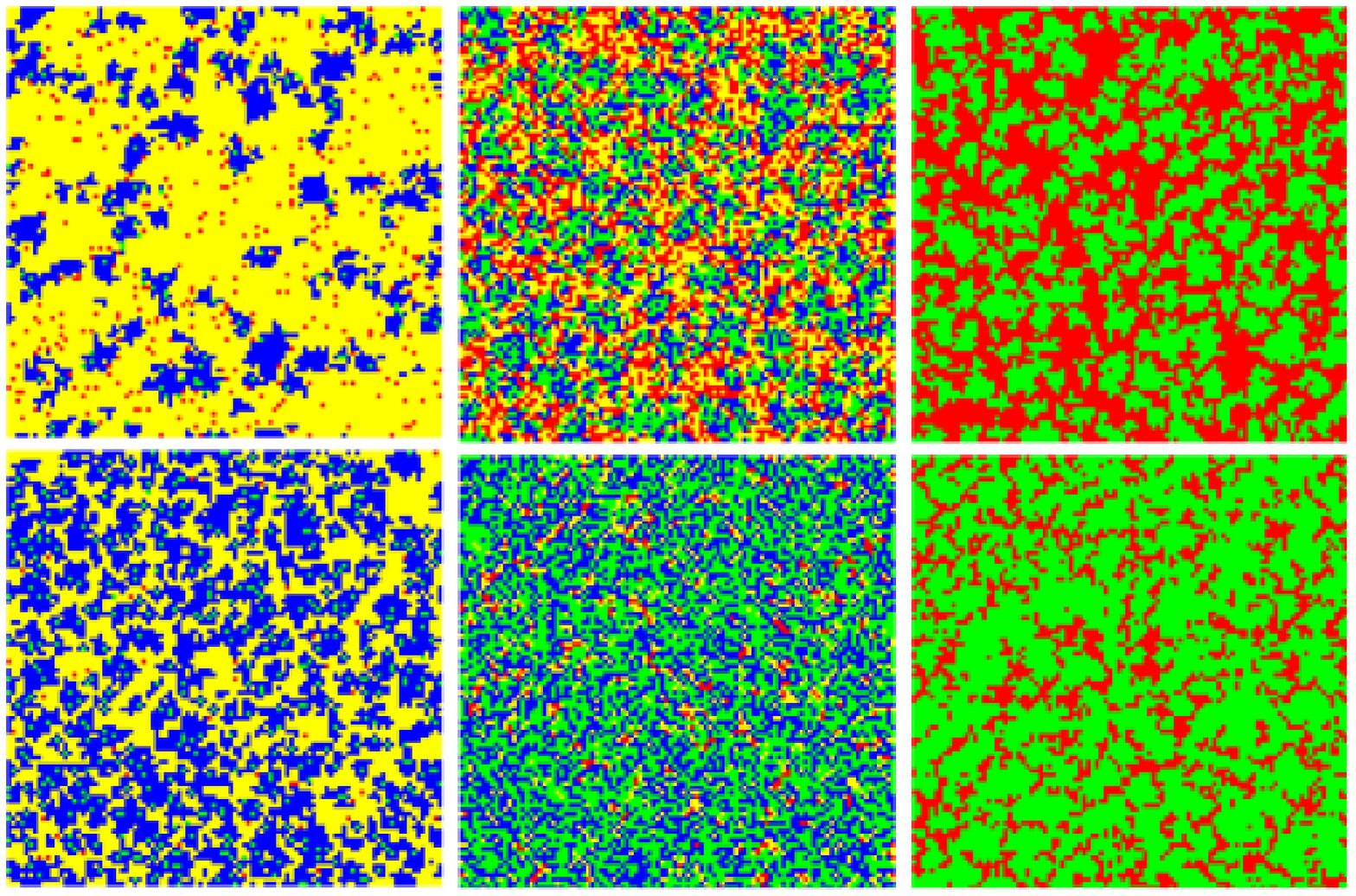}
\caption{\textbf{Characteristic snapshots of strategy distributions on the square lattice.} Top row depicts results for the aspiration parameter $u=0.25$ while the bottom row features results for $u=1.0$. In both rows the fraction of type $A$ players $v$ is $0.05$, $0.5$ and $1.0$ from left to right. Cooperators of type $A$ and $B$ are colored green and blue, respectively. Defectors of type $A$ and $B$, on the other hand, are colored red and yellow. If comparing the snapshots vertically, it can be observed that larger values of $u$ (top $0.25$, bottom $1.0$) clearly promote the evolution of cooperation. The scenario from left to right via increasing the fraction of type $A$ players is not so clear cut. For $u=0.25$ (top row) we can conclude that larger $v$ favor cooperative behavior, as clearly the cooperators flourish more and more from the left toward the right panel. For $u=1.0$ (bottom row), however, it seems that for $v=0.5$ (bottom middle) cooperators actually fare better then for both $v=0.05$ (bottom left) and $v=1.0$ (bottom right). Hence, the conclusion imposes that for higher $u$ values an intermediate (rather than the maximal, as is the case for lower $u$) fraction of type $A$ players (those that aspire to their most fittest neighbors only) is optimal for the evolution of cooperation. Results in all panels were obtained for $r=0.02$ and $K=0.1$.}
\label{fig1}
\end{figure}

It is instructive to first examine characteristic snapshots of the square lattice for different values of $u$ and $v$. Results presented in Fig.~\ref{fig1} hint to the conclusion that heterogenous aspiration to the fittest promotes cooperation, although the details of this claim depend somewhat on the value of the aspiration parameter $u$. For small values of $u$ it is best if all the players, \textit{i.e.} $v=1$, aspire to their slightly (note that $u$ is small) fitter neighbors and thus none actually choose the potential strategy sources uniformly at random. This can be deduced from the top three panels of Fig.~\ref{fig1} if compared from left to right. For large $u$, however, it is best if only half of the players, \textit{i.e.} $v \approx 0.5$, aspire to their most fittest neighbors, while the other half chooses their role models randomly. This can be observed if one compares the bottom three panels of Fig.~\ref{fig1} with one another, although the difference in the overall density of cooperators (depicted green and blue) between the middle and the right panel is fairly small. Finally, the role of the aspiration parameter is more clear cut since larger $u$ clearly favor the cooperative strategy if compared to small $u$. This can be observed if comparing the snapshots presented in Fig.~\ref{fig1} vertically.

\begin{figure}[!ht]
\begin{center}
\includegraphics[width=11.033cm]{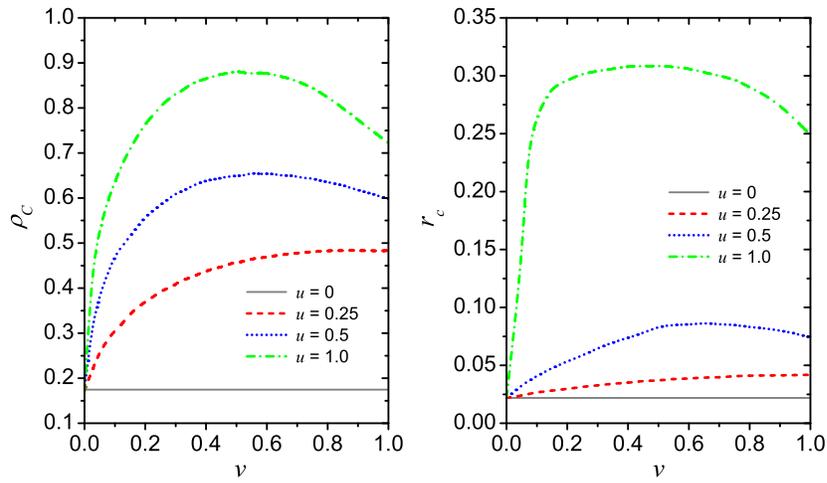}
\end{center}
\caption{\textbf{Promotion of cooperation due to heterogenous aspirations on the square lattice.} Left panel depicts the density of cooperators $\rho_{C}$ in dependence on the fraction of type $A$ players $v$ for different values of the aspiration parameter $u$. Right panel depicts the critical cost-to-benefit ratio $r=r_{c}$ at which cooperators die out, \textit{i.e.} $\rho_{C}=0$, in dependence on $v$ for different values of $u$. Results in both panels convey the message that low values of $u$ require a high fraction of type $A$ players for cooperation to flourish. Conversely, higher values of $u$ sustain cooperation optimally if only half ($v \approx 0.5$) of the players aspires to their most fittest neighbors while the rest chooses whom to potentially imitate uniformly at random. Optimal conditions for the evolution of cooperation thus require $u$ and $v$ to be fine-tuned jointly. Depicted results in both panels were obtained for $r=0.02$ and $K=0.1$.}
\label{fig2}
\end{figure}

Since the snapshots presented in Fig.~\ref{fig1} can be used primarily for an initial qualitative assessment of the impact of heterogeneous aspirations, we present in Fig.~\ref{fig2} the fraction of cooperators $\rho_{C}$ (left) and the critical cost-to-benefit ratio $r_c$ (right) in dependence on $v$ for different values of $u$. It can be observed that the promotion of cooperation for the optimal combination of the two parameters, being $u=1$ and $v \approx 0.5$, is really remarkable. The fraction of cooperators rises from $0.18$ to $0.87$, while the critical cost-to-benefit ratio rises a full order of magnitude from $r_c=0.022$ to $0.31$. As tentatively deduced from the lower three snapshots in Fig.~\ref{fig1}, it can also be observed that for high values of $u$ an intermediate fraction of type $A$ players is optimal for the evolution of cooperation. Conversely, for low $u$ the fraction of cooperators $\rho_{C}$ and the critical cost-to-benefit ratio $r_c$ both increase monotonously with increasing $v$. If, however, selecting a particular value of $v$, then the impact of the aspiration parameter $u$ is always such that cooperation is the more promoted the larger the value of $u$. This can be observed clearly from both panels, and indeed seems like the main driving force behind the elevated levels of cooperation. Fine-tuning the fraction of players making use of the aspiration to the fittest (from $v=1$ downwards since the $v=0$ limit trivially returns the random selection of potential strategy sources) at high $u$ can rise the cooperation level further, but more in the sense of minor adjustments, similarly as was observed in the past for the impact of uncertainty by strategy adoptions \cite{vukov_pre06} or the impact of noise \cite{perc_njp06a}.

\begin{figure}[!ht]
\begin{center}
\includegraphics[width=11.033cm]{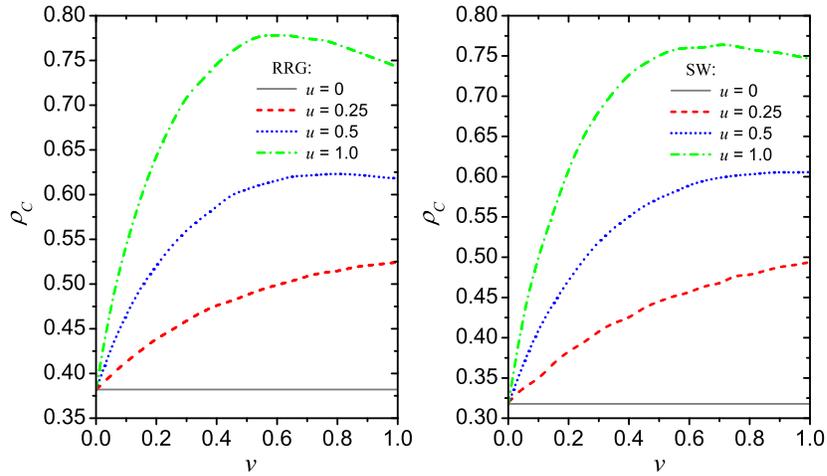}
\end{center}
\caption{\textbf{Promotion of cooperation due to heterogenous aspirations on the random regular graph (RRG) and the small-world (SW) network.} Left panel depicts the density of cooperators $\rho_{C}$ in dependence on the fraction of type $A$ players $v$ for different values of the aspiration parameter $u$ for the RRG. Right panel depicts $\rho_{C}$ in dependence on $v$ for different values of $u$ for the Watts-Strogatz SW network with the fraction of rewired links equalling $0.1$. These results are in agreement with those presented in Fig.~\ref{fig2}, supporting the conclusion that the impact of heterogenous aspirations on the evolution of cooperation is robust against alterations of the interaction network. As on the square lattice, low, but also intermediate, values of $u$ require $v=1.0$ for cooperation to thrive, while higher values of $u$ sustain cooperation optimally only if $v \approx 0.6$. Depicted results in both panels were obtained for $r=0.02$ and $K=0.1$.}
\label{fig3}
\end{figure}

Aiming to generalize the validity of our results, we present in Fig.~\ref{fig3} the fraction of cooperators $\rho_{C}$ in dependence on $v$ for different values of $u$ as obtained on the random regular graph (left) and the small-world network (right). The goal is to test to what extend above conclusions hold also on interaction networks other than the square lattice, in particular such that are more complex and spatially heterogenous. If comparing the obtained results with those presented in the left panel of Fig.~\ref{fig2}, it seems save to conclude that they are to a very large degree qualitatively identical. Some differences nevertheless can be observed. The first is that what constitutes a high $u$ limit is a bit higher on complex networks than on the regular lattice. Note that for $u=0.5$ the optimal fraction of type $A$ players is practically still $v \to 1$. Even for $u=1.0$ the bell-shaped dependence on $v$ is far less pronounced than on the square lattice, and the optimal $v$ (the peak of $\rho_{C}$) is closer to $0.6$ than $0.5$. The second difference is, looking relatively to the starting point at $u=v=0$, that the promotion of cooperation due to positive $u$ and $v$ is somewhat less prolific. This is, however, not that surprising since complex networks in general promote cooperation already on their own \cite{szabo_pr07}, and thus secondary promotive mechanisms may therefore become less expressed. Aside from these fairly mild differences though, we can conclude that heterogenous aspirations do promote cooperation irrespective of the underlying interaction network, and that the details of the promotive effect are largely universal and predictable.

\begin{figure}[!ht]
\includegraphics[width=15.50cm]{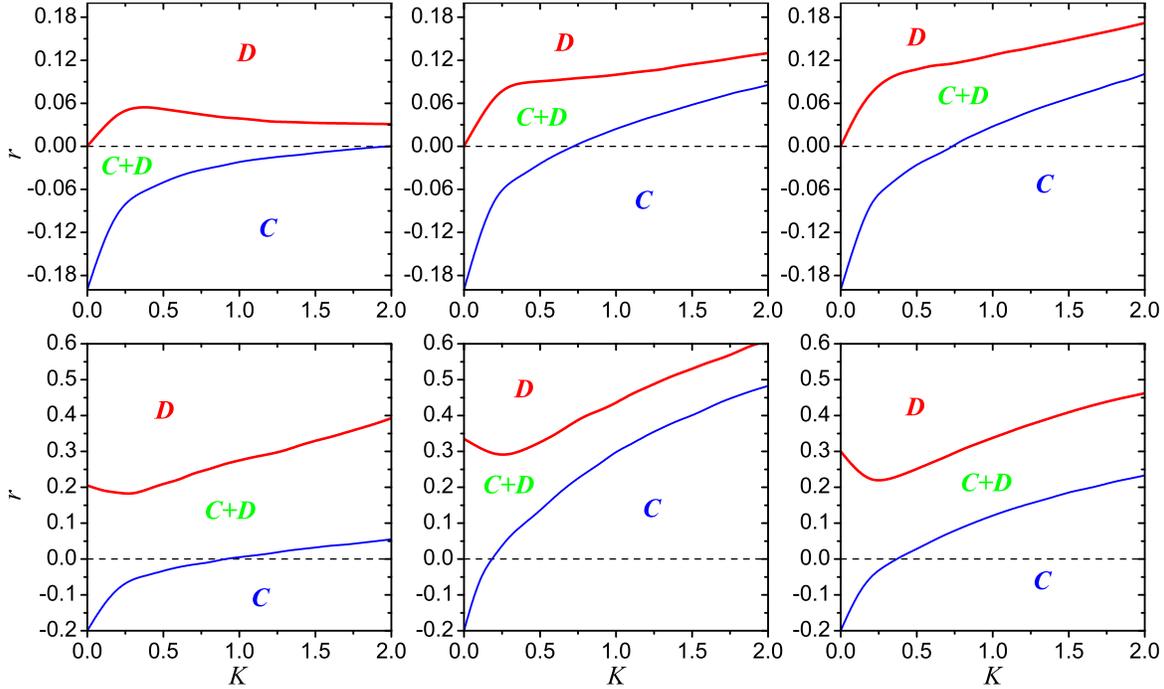}
\caption{\textbf{Full $r-K$ phase diagrams for the square lattice.} Top row depicts results for the aspiration parameter $u=0.25$ while the bottom row features results for $u=1.0$. In both rows the fraction of type $A$ players $v$ is $0.05$, $0.5$ and $1.0$ from left to right. The outline of panels thus corresponds to the snapshots presented in Fig.~\ref{fig1}. Thin blue and thick red lines mark the border between stationary pure $C$ and $D$ phases and the mixed $C+D$ phase, respectively. In agreement with previous works \cite{szabo_pre05, vukov_pre06}, it can be observed that for $u=0.25$ and $v=0.05$ (top left) there exists an intermediate uncertainty in the strategy adoption process (an intermediate value of $K$) for which the survivability of cooperators is optimal, \textit{i.e.} $r_{c}$ is maximal. Conversely, while the borderline separating the pure $C$ and the mixed $C+D$ phase for all the other combinations of $u$ and $v$ exhibits a qualitatively identical outlay as for the $u=0.25$ and $v=0.05$ case, the $D \leftrightarrow C+D$ transition is qualitatively different and very much dependent on the particularities players' aspirations. Note that in all the bottom panels there exist an intermediate value of $K$ for which $r_{c}$ is minimal rather than maximal, while towards the large $K$ limit $r_{c}$ increases, saturating only for $K>4$ (not shown). In the top middle and right panel, on the other hand, the bell-shaped outlay of the $D \leftrightarrow C+D$ transition gives way to a monotonically increasing curve, again saturating only for $K>4$. It can thus be concluded that, while the aspiration based promotion of cooperation is largely independent of $K$, the details of phase transition are very much affected, which can be attributed to an effective alterations of the interaction network due to preferred strategy sources (see also main text for details).}
\label{fig4}
\end{figure}

Next, we proceed with examining how positive values of $u$ and $v$ fare under different levels of uncertainty by strategy adoptions. The latter can be tuned via $K$ [see Eq.~(3)], which acts as a temperature parameter in the employed Fermi strategy adoption function \cite{szabo_pre98}. Accordingly, when $K \to \infty$ all information is lost and the strategies are adopted by means of a coin toss. Note that this aspect has thus far not received any attention here as $K=0.1$ was fixed. The matter is not trivial to address because uncertainty and noise can have a rather profound impact on the evolution of cooperation \cite{perc_njp06a, perc_njp06b, vukov_pre06, ren_pre07, tanimoto_pre07b, traulsen_jtb07}, and thus care needs to be exercised. The safest and most accurate way to approach the problem is by means of phase diagrams. Since we have two additional parameters ($u$ and $v$) against which we want to test the impact of $K$, we determined full $r-K$ phase diagrams for six characteristic combinations of $u$ and $v$ on the square lattice. Obtained results are presented in Fig.~\ref{fig4}. Notably, the phase diagram presented in the top left panel of Fig.~\ref{fig4} is well-known, implying the existence of an optimal level of uncertainty for the evolution of cooperation, as was previously reported in \cite{perc_njp06a, vukov_pre06}. In particular, note that the $D \leftrightarrow C+D$ transition line is bell shaped, indicating that $K \approx 0.38$ is the optimal temperature at which cooperators are able to survive at the highest value of $r$. Importantly though, this phenomenon can only be observed on interaction topologies lacking overlapping triangles \cite{szabo_pre05, szolnoki_pre09c}. Interestingly, increasing $u$ from $0.25$ (top row) to $1.0$ (bottom row) completely eradicates (as do interaction networks incorporating overlapping triangles) the existence of an optimal $K$, and in fact qualitatively reverses the dependence. The $D \leftrightarrow C+D$ transition line has an inverted bell-shaped outlay, indicating the existence of the worst rather than an optimal temperature $K$ for the evolution of cooperation. The qualitative changes are less profound if $u$ is kept constant at $0.25$ (top row) and $v$ increases (from left to right). Still, however, the bell-shaped outlay of the $D \leftrightarrow C+D$ transition gives way to a monotonically increasing curve, saturating only for high $K$. These qualitative changes in the phase diagrams imply that increasing the aspiration parameter $u$ or the fraction of players abiding to it (type $A$) effectively alters the interaction network. While the square lattice obviously lacks overlapping triangles and thus enables the observation of an optimal $K$ for small enough values of $u$ and $v$ (or a combination thereof, as is the case in the top left panels), trimming the likelihood of who will act as a strategy source and how many players will actually aspire to their fittest neighbors seems to effectively enhance linkage among essentially disconnected triplets and thus precludes the same observation. It is instructive to note that a similar phenomenon was observed recently in public goods games, where the joint membership in large groups was also found to alter the effective interaction network and thus the impact of uncertainly on the evolution of cooperation \cite{szolnoki_pre09c}.

In terms of the facilitation of cooperation, however, it can be concluded that the promotive impact of positive values of $u$ and $v$ prevails irrespective of $K$. By comparing the extend of pure $C$ and mixed $C+D$ regions for different pairs of the two parameters, we can observe that for small values of $u$ (top panels in Fig.~\ref{fig4}) it is best if all the players, \textit{i.e.} $v=1$, aspire to their slightly (note that $u$ is small) fitter neighbors, while for large $u$ (bottom panels in Fig.~\ref{fig4}) it is best if only approximately half of the players, \textit{i.e.} $v \approx 0.5$, aspire to their most fittest neighbors. The same conclusions were stated already upon the inspection of results presented in Figs.~\ref{fig2} and \ref{fig3}, and with this we now affirm that not only is the promotion of cooperation via heterogeneous aspirations robust against differences in the interaction networks, but also against variations in the uncertainty by strategy adoptions.

\begin{figure}[!ht]
\begin{center}
\includegraphics[width=7.415cm]{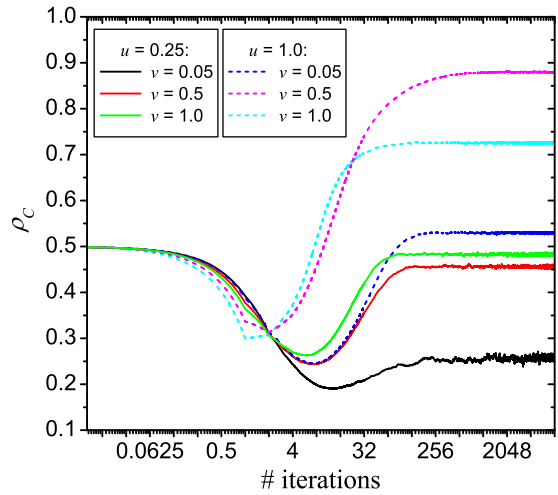}
\end{center}
\caption{\textbf{Time courses of the density of cooperators on the square lattice.} Results are presented for the aspiration parameters $u=0.25$ (solid lines) and $u=1.0$ (dashed lines), each for three different fractions of type $A$ players $v$, as depicted on the figure. The crucial feature of all time courses is the initial temporary downfall of cooperators, which sets in for all depicted combinations of $u$ and $v$. Quite remarkably, what appears to become an ever faster extinction eventually becomes a rise to, at least in some cases, near-dominance. Note that the horizontal axis is logarithmic and that values of $\rho_{C}$ were recorded also in-between full iteration steps to ensure a proper resolution. Depicted results were obtained for $r=0.02$ and $K=0.1$.}
\label{fig5}
\end{figure}

It remains of interest to elucidate why then cooperative behavior is in fact promoted by positive values of $u$ and $v$. To provide answers, we show in Fig.~\ref{fig5} time courses of $\rho_{C}$ for different characteristic combination of the two main parameters that we have used throughout this work. What should attract the attention is the fact that in the most early stages of the evolutionary process (note that values of $\rho_{C}$ were recorded also in-between full iteration steps) it appears as if defectors would actually fare better than cooperators. This is actually what one would expect, given that defectors are, as individuals, more successful than cooperators and will thus be chosen more likely as potential strategy donors if $u$ is positive. This should in turn amplify their chances of spreading and ultimately result in the decimation of cooperators (indeed, only between 20-30~\% survive). Quite surprisingly though, the tide changes fairly fast, and as one can observe from the presented time courses, frequently the more so the deeper the initial downfall of cooperators. We argue that for positive values of $u$ and $v$ a negative feedback effect occurs, which halts and eventually reverts what appears to be a march of defectors toward dominance. Namely, in the very early stages of the game defectors are able to plunder very efficiently, which quickly results in a state where there are hardly any cooperators left to exploit. Consequently, the few remaining clusters of cooperators start recovering lost ground against weakened defectors. Crucial thereby is the fact that the clusters formed by cooperators are impervious to defector attacks even at high values of $r$ because of the positive selection towards the fittest neighbors acting as strategy sources (occurring for $u>0$). In a sea of cooperators this is practically always another cooperator rather than a defector trying to penetrate into the cluster. This newly identified mechanism ultimately results in fairly widespread cooperation that goes beyond what can be warranted by the spatial reciprocity alone (see \textit{e.g.} \cite{szabo_pr07}), and this irrespective of the underlying interaction network and uncertainty by strategy adoptions.

\begin{figure}[!ht]
\begin{center}
\includegraphics[width=7.415cm]{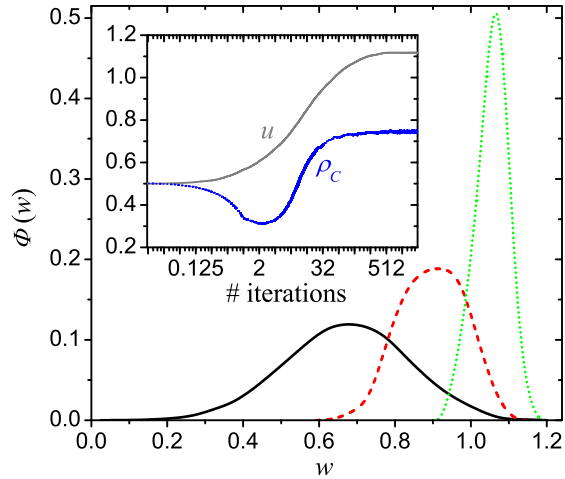}
\end{center}
\caption{\textbf{Spontaneous fixation towards an intermediate aspiration level by means of natural selection.} Presented results were obtained with the alternative model where players are not divided into two groups and initially every player is assigned a random aspiration level $w_{x}$ drawn from a Gaussian distribution with the mean $\mu=0.5$ and standard deviation $\sigma=0.167$. The main panel depicts the distributions $\Phi(w)$ of individual aspiration levels as recorded at $4$ (solid black line), $32$ (dashed red line) and $256$ (dotted green line) full iteration steps. The fixation towards a dominant average value $u=L^{-2}\sum_x w_x$ due to natural selection is evident since the interval of $w$ values still present in the population becomes more and more narrow as time progresses. The inset shows the convergence of $u$ (solid gray line) and $\rho_{C}$ (dotted blue line). The initial temporary downfall of cooperators, followed by the rise to near-dominance, is well-expressed also in the coevolutionary setup, and the stationary density agrees well with the results obtained by means of the original model with $v=1.0$ and $u=1.0$ (compare with the dashed cyan line in Fig.~\ref{fig5}). Note that in the inset the horizontal axis is logarithmic and that values of $u$ and $\rho_{C}$ were recorded also in-between full iteration steps to ensure a proper resolution. Depicted results were obtained for $r=0.02$ and $K=0.1$.}
\label{fig6}
\end{figure}

Finally, it is instructive to examine whether an optimal intermediate value of $w_{x}$, determining the aspiration level of player $x$, can emerge spontaneously from an initial array of normally distributed values. This would imply that natural selection indeed favors individuals with a specific aspiration level, which would in turn extend the credibility of thus far presented results that were obtained primarily in a top-down manner [by optimizing a population-level property (cooperation) by means of an appropriate selection of parameters determining the aspiration level of individuals]. For this purpose we omit the division of the population on players of type $A$ and $B$, and initially assign to every player a value $w_{x}$ that is drawn randomly from a Gaussian distribution with a given mean $\mu$ and standard deviation $\sigma$. Then if player $x$ adopts the strategy from player $y$ also $w_{x}$ becomes equal to $w_{y}$ (see Methods for details). Results obtained with this alternative coevolutionary model are presented in Fig.~\ref{fig6}. It can be observed that the initial Gaussian distribution sharpens fast around an intermediate value of $w$, which then gradually becomes more and more frequent in the population as the natural selection spontaneously eliminates the less favorable values that warrant a lower individual fitness. The final state is a population where virtually all players have an identical aspiration level $w_{x}=w$, and accordingly, the outcome in terms of the stationary density of cooperators is equal to that obtained with the original model having $v=1$ and $u=w$. In this sense the preceding results are validated and their generality extended by means of a bottom-up approach entailing a spontaneous coevolution towards an intermediate individually optimal aspiration level. We note, however, that with this simple coevolutionary model the result that heterogeneous aspirations promote cooperation is not exactly reproduced. Further studies on more sophisticated models incorporating coevolving aspirations are required to arrive spontaneously at a heterogeneous distribution of individual aspiration levels. Inspirations for this can be found in the recent review on coevolutionary games \cite{perc_bs10}, and we are looking forward to further developments in this direction.

\section*{Discussion}
We have shown that heterogenous aspiration to the fittest, \textit{i.e.} the propensity of designating the most successful neighbor as being the role model, may be seen as a universally applicable promoter of cooperation that works on different interaction networks and under different levels of stochasticity. For low and moderate values of the aspiration parameter $u$ cooperation thrives best if the total population abides to aspiring to the fittest. For large values of $u$, however, it is best if only approximately half of the players persuasively attempt to copy their most successful neighbors while the rest chooses their opponents uniformly at random. The optimal evolution of cooperation thus requires fine-tuning of both, the density of players that are prone to aspiring to the fittest, as well as the aspiration parameter that determines how fit a neighbor actually must be in order to be considered as the potential source of the new strategy. In addition, by studying an alternative model where individual aspiration levels were also subject to evolution, we have shown that an intermediate value of the aspiration level emerges spontaneously through natural selection, thus supplementing the main results by means of a coevolutionary approach.

Notably, the extensions of the prisoner's dilemma game we have considered here seem very reasonable and are in fact easily justifiable with realistic examples. For example, it is a fact that people will, in general, much more likely follow a successful individual than somebody who is just struggling to get by. Under certain adverse circumstances, like in a state of rebelion or in revolutionary times, however, it is also possible that individuals will be inspired to copy their less successful partners or those that seem to do more harm than good. In many ways it seems that the ones who are satisfied with just picking somebody randomly to aspire to are the ones that are most difficult to come by. In this sense the rather frequently adopted random selection of a neighbor, retrieved in our case if $u=0$ (or equivalently $v=0$), seems in many ways like the least probable alternative. In this sense it is interesting to note that our aspiring to the fittest becomes identical to the frequently adopted, especially in the early seminal works on games on grids \cite{nowak_n92b, nowak_ijbc93, nowak_ijbc94}, ``best takes all'' adoption rule if $v=1$, $u \to \infty$ in Eq.~(2), and $K \to 0$ in Eq.~(3). Although in our simulations we never quite reach the ``best takes all'' limit, and thus a direct comparison with the seminal works is somewhat circumstantial, we find here that the intermediate regions of heterogenous aspirations offer fascinating new insights into the evolution of cooperation, and we hope that this work will inspire future studies, especially in terms of understanding the emergence of successful leaders in societies via a coevolutionary process \cite{perc_bs10}.

\section*{Methods}
An evolutionary prisoner's dilemma game with the temptation to defect $T = b$ (the highest payoff received by a defector if playing against a cooperator), reward for mutual cooperation $R = b-c$, the punishment for mutual defection $P=0$, and the sucker's payoff $S=-c$ (the lowest payoff received by a cooperator if playing against a defector) is used as the basis for our simulations. Without loss of generality the payoffs can be rescaled as $R=1$, $T=1+r$, $S=-r$ and $P=0$, where $r=c/(b-c)$ is the cost-to-benefit ratio \cite{hauert_ajp05}. For positive $r$ we have $T>R>P>S$, thus strictly satisfying the prisoner's dilemma payoff ranking.

As the interaction network, we use either a regular $L \times L$ square lattice, the random regular graph (RRG) constructed as described in \cite{szabo_jpa04}, or the small-world (SW) topology with an average degree of four generated via the Watts-Strogatz algorithm \cite{watts_d_n98}. Each vertex $x$ is initially designated as hosting either players of type $n_x=A$ or $B$ with the probability $v$ and $1-v$, respectively. This division of players is performed uniformly at random irrespective of their initial strategies and remains unchanged during the simulations. According to established procedures, each player is initially also designated either as a cooperator ($s_x=C$) or defector ($D$) with equal probability. The game is iterated forward in accordance with the sequential simulation procedure comprising the following elementary steps. First, player $x$ acquires its payoff $p_x$ by playing the game with all its neighbors. Next, we evaluate in the same way the payoffs of all the neighbors of player $x$ and subsequently select one neighbor $y$ via the probability
\begin{equation}
\Pi_{y}=\frac{\exp(w_{y} p_{y})}{\sum_{z} \exp(w_{z} p_{z})},
\end{equation}
where the sum runs over all the neighbors of player $x$. Importantly, $w_{x}$ is the so-called selection or aspiration parameter that depends on the type of player $x$ according to \begin{equation}
w_{x}=\left\{ \begin{array}{ll}
u, & \textrm{if $n_{x}=A$}\\
0, & \textrm{if $n_{x}=B$.}
\end{array} \right.
\end{equation}
Evidently, if the aspiration parameter $u=0$ then irrespective of $v$ (density of type $A$ players) the most frequently adopted situation is recovered where player $y$ is chosen uniformly at random from all the neighbors of player $x$. For $u>0$ and $v>0$, however, Eqs.~(1) and (2) introduce a preference in all players of type $A$ (but not in players of type $B$) to copy the strategy of those neighbors who have a high fitness, or equivalently, a high payoff $p_{y}$. Lastly then, after the neighbor $y$ that is aspired to by player $x$ is chosen, player $x$ adopts the strategy $s_y$ from the selected player $y$ with the probability
\begin{equation}
W(s_y \rightarrow s_x)=\frac{1}{1+\exp[(p_x-p_y)/K]},
\end{equation}
where $K$ denotes the amplitude of noise or its inverse ($1/K$) the so-called intensity of selection \cite{szabo_pre98}. Irrespective of the values of $u$ and $v$ one full iteration step involves all players $x=1,2, \ldots, L^2$ having a chance to adopt a strategy from one of their neighbors once.

An alternative model, allowing for individual $w_{x}$ values to be subject to evolution as well, entails omitting the division of the population on two types of players and assigning to every individual an initial $w_{x}$ value that is drown randomly from a Gaussian distribution having mean $\mu$ and standard deviation $\sigma$, as was done recently in \cite{mcnamara_n08}, for example. Subsequently, if player $x$ adopts the strategy from player $y$ following the identical procedure as described above for the original model, then the value of $w_{x}$ changes to that of $w_{y}$ as well. The key question that we aim to answer with this model is whether a specific aspiration level is indeed optimal for an individual to prosper, and if yes, does the selection pressure favor it spontaneously. Essentially, we are interested in the distribution of $w_{x}$ values after the stationary fraction of strategies in the population is reached. A link with the original model can be established by considering in this case $v$ to equal one and $u=L^{-2}\sum_x w_x$.

Results of computer simulations were obtained on populations comprising $100 \times 100$ to $400 \times 400$ individuals, whereby the fraction of cooperators $\rho_{C}$ was determined within $10^5$ full iteration steps after sufficiently long transients were discarded. Moreover, since the heterogeneous preferential selection of neighbors may introduce additional disturbances, final results were averaged over up to $40$ independent runs for each set of parameter values in order to assure suitable accuracy.

\section*{Acknowledgments}
We have benefited substantially from the insightful comments of the Editor James Arthur Robert Marshall and an anonymous referee, which were instrumental for elevating the quality of this work. Helpful discussions with Professor Lianzhong Zhang are gratefully acknowledged as well.

\clearpage


\end{document}